\documentstyle[aps,prb,epsf]{revtex} 
\begin{document}
\title{ Coulomb interaction at Superconductor to Mott-insulator transition}
\author{Jinwu Ye}
\address{Department of Physics and Astronomy,
The Johns Hopkins University, Baltimore MD, 21218}
\date{\today}
\maketitle
\begin{abstract}
We reexamine the effects of long-range Coulomb interaction
on the onset of superconductivity. We use the model of $ N $
complex scalar fields with the Coulomb interaction studied first
by Fisher and Grinstein (FG).
We find that near d=3 space dimension, the system
undergoes second order phase transition if $ N \ge 55.39 $,
but undergoes possible fluctuation driven first order transition
if $ N< 55.39 $. We give the detailed derivation of the field theory
Renormalization Group (RG) of this model to one loop. Our RG results disagree
with those of FG near $ d=3 $. Possible scenario at $ d=2 $
is proposed.
\end{abstract}

\section{Introduction}

  Superconductor to insulator (SI) transitions have been observed 
  in disordered thin
  films by systematically varying the thickness of the films \cite{film}
  or by tuning the magnetic fields \cite{mag}. A closely related SI
  transition has also been studied in
  artificially constructed 2d Josephson-junction arrays by systematically
  varying the ratio of charging energy and Josephson coupling
  energy \cite{junction} or by tuning the magnetic fields \cite{junction1}.
  Universal conductivities have been measured right at the SI transitions
  in these systems.
  More recently, SI transition was demonstrated in 3d granular Al-Ge
  samples \cite{3d}.
  Many theoretical efforts \cite{z=1,coul,glass,magtune,kos,wen,cha} 
  have been made to investigate the SI transitions
  in these experimental systems. 
  It is generally accepted that the transition is correctly described by 
  a model of {\em interacting} charge-2e bosons moving in a 2d random potential.
  A related, but much simpler model is Boson-Hubbard model which
  consists of bosons hopping on a periodic lattice.
  When the boson density is commensurate with
  the lattice, this model displays a quantum transition from Mott insulator to a
  superconductor.
  Neglecting disorder as a first approximation, a lot of authors
  studied the SI transition and its universal conductivity
  in this simplified model \cite{coul,wen,cha,subir}.
  Near the quantum critical point, the effective low energy 
  theory of this model is simply a $ \phi^{4} $ theory:
\begin{equation}
   {\cal S}  =  \int d^d x d \tau \Biggl[ 
   | \partial_{0} \phi_{m} |^{2} + | \partial_{i} \phi_{m} |^{2} +
   r |\phi_{m} |^{2} + \frac{u}{4} |\phi_{m} |^{4} \Biggr]
\end{equation}
    where $\phi_m$ are $m=1\cdots N$ species of charge
    $ e^{*}= 2e $ complex bosons.

    In order to confront with experiments, several important effects such as
    disorder, dissipation and long-range Coulomb interaction should be 
    taken into account. From general scaling arguments, Fisher, Grinstein and
    Girvin argued that the long range Coulomb interaction
    between the charged
    bosons will render the dynamic exponent $ z $ to be 1 in any dimension
    \cite{z=1}.
    This is indeed what was observed on the magnetic field
    tuned SI transition in thin 2D films \cite{mag}.
    The measurements on magnetic field-induced SI transitions in
    Josephson-Junction arrays are also consistent with $ z=1, \nu>1$.
    Therefore, it is very important to investigate the effects of
    Coulomb interaction on SI transition in any detail.
    In Ref.\cite{coul}, Fisher and Grinstein
     (FG) studied the model of charged bosons with Coulomb interaction
    hopping on a lattice. By taking continuum limit, they found that the
    low energy effective theory is described by
\begin{equation}
   {\cal S}  =  \int d^d x d \tau \Biggl[ 
   | ( \partial_{0}-i e^{*} A_{0}) \phi_{m} |^{2} + | \partial_{i} \phi_{m} |^{2}
   + r |\phi_{m} |^{2} + \frac{u}{4} |\phi_{m} |^{4} \Biggr ] +
     \frac{1}{2} \int \frac{ d^d k}{ (2 \pi)^{d} } \frac{ d \omega}{ 2 \pi}
      k^{\sigma} | A_{0} ( \vec{k}, \omega ) |^{2}
\label{fg}
\end{equation}
     Where $ A_{0} $ is the time component of the $ U(1) $ gauge fields
     and is responsible for the long-range interaction between the
     charged bosons. The $ 1/r $ Coulomb interaction corresponds to
     $ \sigma=d-1 $.
     
   The model Eq.\ref{fg} should not be confused with the model
   studied in Refs.\cite{bert,chen,hi}. The crucial
   difference is that only the time component $ A_{0} $ of gauge
   fields is involved in Eq.\ref{fg}, therefore, the Lorentz invariance
   is broken. However, in the scalar electrodynamics studied in
   Refs.\cite{bert,chen,hi},
   both time and transverse components are involved, the Lorentz
   invariance is respected.
     
    In principle, the coupling of the order parameter to the spatial
    components of the $ U(1) $ gauge fields should also be included
    in the above equation. However, at 2D films, the penetration depth
    $\lambda_{2d} = \frac{\lambda^{2}}{d} $ with $ \lambda $ the {\em bulk}
    penetration depth and $ d $ the film thickness. In the experimental
    systems \cite{film}, $\lambda \sim 100 \stackrel{\circ}{A},
    d \sim 5 \stackrel{\circ}{A} $,
    so $ \lambda_{2d} \sim 2000 \stackrel{\circ}{A} $, the spatial coupling
    will not manifest itself until experimentally unobservably close to the 
    SI transition \cite{coul,magtune,length}. 
    Therefore, it is safe to neglect the spatial coupling
    in Eq.\ref{fg}. In 3D samples investigated in Ref.\cite{3d}, we assume
    that they are in extreme type-II limit. In this limit, the spatial
    coupling can also be neglected.

     FG  did Renormalization Group (RG) analysis
     of the model by performing a double expansion in $ \epsilon=3-d $
     and $ \epsilon_{\sigma}=2-\sigma $. The $ 1/r $ Coulomb interaction
     corresponds to $ \epsilon=\epsilon_{\sigma} $. They
     concluded that depending on parameters, at two
     space dimension, the transition can be either first order or 
     second order with Coulomb coupling marginally irrelevant at
     the 3D XY critical point.

      In this paper, we provide a detailed RG analysis of the
  Model Eq.\ref{fg} using field theory method.
  We do not perform a double expansion in  $ \epsilon $
  and $ \epsilon_{\sigma} $. Instead, for simplicity,
  we fix $ \sigma  $ to be 2 ( namely $ \epsilon_{\sigma} \equiv 0 $ )
  and perform a $ \epsilon= 3-d $
  expansion. $ \sigma \equiv 2 $ corresponds to $ 1/r $ Coulomb interaction
  at $ d=3 $ and {\em logarithmic } instead of $ 1/r $ interaction at $ d=2 $.
We find that {\em near} 3 space dimension, the system
undergoes second order phase transition if $ N \ge 55.39 $,
but undergoes {\em possible} fluctuation-driven first order transition
if $ N<55.39 $. We calculate the dynamic exponent $ z $, the
boson propagator exponent $ \eta $, the gauge field ( Coulomb interaction )
exponent $ \eta_{A} $ and the correlation exponent $ \nu $.  
We give the detailed derivation of field theory
RG of this model, because the RG method we used
is interesting in its own right. We expect the structure brought
out by our RG procedures have some general impacts on 
how to perform correct RG on other zero temperature
quantum critical phenomena \cite{hertz}.
 The same method has been applied successfully to study the quantum
 transition from Fractional Quantum Hall state to insulating state
 in a periodic potential \cite{fermion}.

  If putting $ \epsilon_{\sigma} $ to zero in Eq. (7) of Ref.\cite{coul},
  we find our RG results disagree with those of FG. Although
  no details are given in Ref.\cite{coul},
  we suspect that (1) the anisotropy between space and time
      (namely the dynamic exponent z )
      was not treated correctly by FG (2) The Ward identity (
      Eq.\ref{ward} in Sec. II ) is violated in the momentum shell
      method employed by FG.
  
  In this paper, we do not even intend to perform the double expansion in 
  $ \epsilon=3-d $ and $\epsilon_{\sigma}=2-\sigma $. Because
  we believe that when {\em gauge field} fluctuations are involved,
   extrapolating to the physical case $ \epsilon=\epsilon_{\sigma}=1 $
   can only lead to misleading results.

    The rest of the paper is organized as follows. In Sec. II, we introduce
    the model and set up the formulation to perform RG. In Sec.III, we
    give the detailed RG analysis and bring out its elegant structure. We
    find that our results disagree with those of FG and point out the
    possible mistakes made by FG. In Sec.IV, we give 
    more discussions and point out the possible scenario at $ d=2 $ and
    some future directions. Finally,
    in the appendix, we perform RG on the well-studied Scalar Electrodynamics
    in Feymann gauge as a non-trivial check on the correctness of our
    calculations in Sec.III.

\section{ The Model and Formulation} 

   In order to perform RG analysis, we rewrite Eq.\ref{fg} in the following
   form \cite{omit}
\begin{eqnarray}
   {\cal S} & = & \int d^d x d \tau \Biggl[ \alpha^{2}
   | \partial_{0} \phi_{m} |^{2} + | \partial_{i} \phi_{m} |^{2}
   + r |\phi_{m} |^{2} + \frac{u}{4} \mu^{\epsilon} \alpha
   |\phi_{m} |^{4}  \nonumber  \\
   & - & i e  \mu^{\epsilon/2} \alpha^{3/2} A_{0}(
      \partial_{0} \phi_{m}^{\dagger} \phi_{m}-
      \phi_{m}^{\dagger} \partial_{0} \phi_{m})
      +e^{2} \mu^{\epsilon} \alpha A^{2}_{0}
      \phi_{m}^{\dagger} \phi_{m}+ \frac{1}{2} (\partial_{i} A_{0} )^{2}
	\Biggr]
\label{begin}
\end{eqnarray}

  Where $x_i$ ($\tau$) are spatial (temporal) co-ordinates with
  $\partial_0 \equiv \partial_{\tau}$, $\partial_i \equiv \partial_{x_i}$.
we are working in $d=3+\epsilon$
spatial dimensions and $\mu$ is a renormalization scale. 
The parameter $\alpha$ is introduced to allow for anisotropic renormalization
between space and time~\cite{bre,cardy}.
 
   It is easy to see Eq.\ref{begin} is invariant under the
   {\em space-independent} gauge-transformation
\begin{equation}
 \phi \rightarrow \phi e^{i \Lambda (\tau) },~~~~
 A_{0} \rightarrow A_{0}- \frac{\sqrt{\alpha}}{e}
 \partial_{0} \Lambda (\tau)
\label{tran}
\end{equation}

 In the perturbation theory, the three kinds of vertices
 in Fig.1 are needed

\begin{figure}
\epsfxsize=12 cm
\centerline{\epsffile{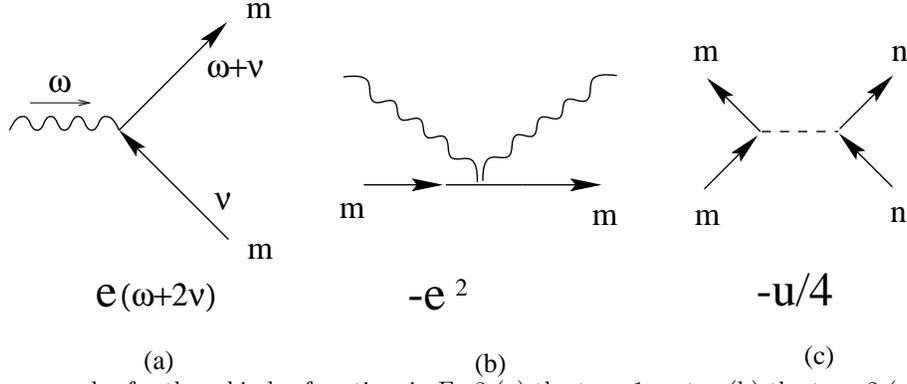}}
\caption{ The Feymann rules for three kinds of vertices in Eq.\ref{begin}
(a) the type 1 vertex (b) the type 2 (seagull) vertex (c)
the quartic coupling of bosons }
\label{vertex}
\end{figure}

The loop expansion requires counterterms to account for ultraviolet divergences
in momentum integrals; we write the counter terms as
\begin{eqnarray}
   {\cal L}_{CT} & = & \alpha^{2}(Z_{\alpha}-1)
   | \partial_{0} \phi_{m} |^{2} +(Z_{2}-1)| \partial_{i} \phi_{m} |^{2}
   + \frac{u}{4} (Z_{4}-1) \mu^{\epsilon} \alpha
   |\phi_{m} |^{4}  \nonumber  \\
   & - & i (Z_{1}-1) e  \mu^{\epsilon/2} \alpha^{3/2} A_{0}(
   \partial_{0} \phi_{m}^{\dagger} \phi_{m}-
   \phi_{m}^{\dagger} \partial_{0} \phi_{m})
   + (Z_{1}-1) e^{2} \mu^{\epsilon} \alpha A^{2}_{0}
   \phi_{m}^{\dagger} \phi_{m}    \nonumber   \\
   & + & \frac{1}{2} (Z_{3}-1) (\partial_{i} A_{0} )^{2}
\label{counter}
\end{eqnarray}

  The Ward identity following from the gauge invariance dictates
\begin{equation}
     Z_{1} = Z_{\alpha}
\label{ward}
\end{equation}

   Using the identity, we relate the bare fields and couplings in
   ${\cal S}$ to the renormalized quantities by 
\begin{eqnarray}
    \phi_{mB} & = & Z_2^{1/2} \phi_m     \nonumber   \\
    A_{0B} & = & Z_3^{1/2} A_0     \nonumber   \\
    \alpha_B & = & (Z_{\alpha}/Z_2 )^{1/2} \alpha    \nonumber   \\
    e_B & = & e \mu^{\epsilon/2} (Z_{\alpha} / Z_2 )^{1/4} Z_{3}^{-1/2}
		        		   \nonumber  \\
    u_{B} & = & u \mu^{\epsilon} Z_{4} Z_{2}^{-3/2} Z_{\alpha}^{-1/2} 
\label{conn}
\end{eqnarray}
    
The dynamic critical exponent, $z$ is related to the renormalization
of $\alpha$ by~\cite{cardy}
\begin{equation}
z = 1 - \mu \frac{d}{d\mu} \ln \alpha = 1 - 
   \frac{1}{2} \mu \frac{d}{d\mu} \ln \frac{Z_2}{Z_{\alpha}}
\label{dynamic}
\end{equation}

  We define the exponent $ \eta $ by the {\em equal-time}
  correlation function  
\begin{equation}
  < \phi(x,\tau) \phi^{\dagger}(0,\tau) > \sim x^{-(d+z-2+\eta)}
\end{equation}

   It is easy to see this correlation function is invariant under
 the gauge transformation Eq.\ref{tran}, therefore, $ \eta $ is
 indeed a gauge invariant quantity and given by
\begin{equation}
  \eta=  \mu \frac{d}{d \mu} \ln Z_{2}
\label{eta}
\end{equation}

   The anomalous dimension $ \eta_{A} $ of the gauge field is also a gauge
   invariant quantity:
\begin{equation}
  \eta_{A} =  \mu \frac{d}{d \mu} \ln Z_{3}
\label{etaA}
\end{equation}

    Near the critical point, the gauge field propagator is
\begin{equation}
  < A_{0} (-\vec{k},-\omega) A_{0} (\vec{k},\omega) > \sim 
  \frac{1}{ k^{2-\eta_{A}} }
\end{equation}

    Physically, it means that at the critical point, at $ d $ spatial dimension,
 the long-range interaction takes the form $ r^{-(d-2+\eta_{A}) } $.

Finally, the critical exponent $\nu$ is related to the anomalous dimension
of the composite operator $ \phi^{\dagger}_{m} \phi_{m} $ by
\begin{equation}
\nu^{-1} - 2 = \mu \frac{d}{d \mu} \ln Z_{\phi^{\dagger} \phi}
\end{equation}

   The renormalization constant $Z_{ \phi^{\dagger} \phi}$ can be
   calculated by inserting the operator into the boson
   self-energy diagrams.

\section{ The perturbative RG calculation at one-loop}

\subsection{ The boson self-energy}

   First, we consider the boson self-energy diagrams in Fig.2
\begin{figure}
\epsfxsize=12 cm
\centerline{\epsffile{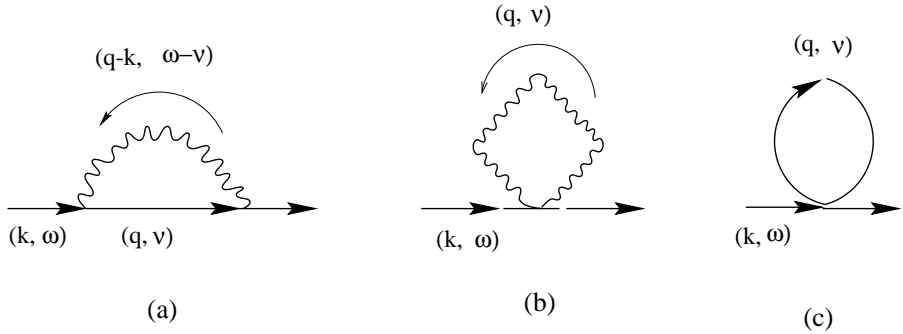}}
\caption{ The self-energy diagram of bosons }
\label{self}
\end{figure}

     Using the Feymann rules listed in Fig.\ref{vertex}, we can bring out
     the divergent structure of Fig.2a by conventional Dimensional
     Regulization (DR) method

\begin{eqnarray}
  (2a) & = & e^{2} \mu^{\epsilon}
  \int \frac{ d^{d} \vec{q}}{ (2 \pi)^{d} } \int \frac{ d \nu}{ 2 \pi}
  \frac{ (\nu + \omega )^{2} }{ (\vec{q}- \vec{k})^{2}(q^{2}+ \nu^{2} ) }
				 \nonumber  \\
      &= & e^{2} \mu^{\epsilon} \int \frac{ d^{d} \vec{q}}{ (2 \pi)^{d} }
  \frac{ 1 }{ (\vec{q}- \vec{k})^{2} } \int \frac{ d \nu}{ 2 \pi}
	 + e^{2} \mu^{\epsilon} \int \frac{ d^{d} \vec{q}}{ (2 \pi)^{d} }
  \frac{ \omega^{2}-q^{2} }{2 q (\vec{q}- \vec{k})^{2} }
				  \nonumber   \\
      &= & e^{2} \mu^{\epsilon} \int \frac{ d^{d} \vec{q}}{ (2 \pi)^{d} }
  \frac{ 1 }{ (\vec{q}- \vec{k})^{2} } \int \frac{ d \nu}{ 2 \pi}
	  + \frac{ e^{2} }{ 4 \pi^{2} \epsilon }
	  ( \omega^{2} -\frac{k^{2}}{3} ) + \cdots
\label{self1}
\end{eqnarray}
    Where $ \cdots $ means all the finite terms.
 
    In the above equation, we scaled all the frequencies by the anisotropy
    parameter $ \alpha $, so $\alpha $ does not appear explicitly in the
    above and in the following.

    Similarly, we can evaluate Fig.2b and 2c
\begin{eqnarray}
  (2b) &= & - e^{2} \mu^{\epsilon} \int \frac{ d^{d} \vec{q}}{ (2 \pi)^{d} }
  \frac{ 1 }{ q^{2} } \int \frac{ d \nu}{ 2 \pi}
		     \nonumber  \\
  (2c) &= & - \frac{ u}{4} \mu^{\epsilon}
   \int \frac{ d^{d} \vec{q}}{ (2 \pi)^{d} }
  \frac{ 1 }{ 2 q }=0
\label{self2}
\end{eqnarray}

   In the last equation, the convention of DR is used. 

  Adding all the contributions from Fig.\ref{self}, we get
\begin{equation}
   (2a)+(2b)+(2c)= \frac{ e^{2} }{ 4 \pi^{2} \epsilon }
	  ( \omega^{2} -\frac{k^{2}}{3} ) + \cdots
\label{add1}
\end{equation}

    It is easy to see that we encounter the UV frequency divergences in both
    Fig.2a and Fig.2b. However, they carry opposite signs, therefore,
    cancel each other. Similar cancellations also appear in the
    Dirac fermion model studied in Ref.\cite{fermion} and will appear
    in the following calculations of other Feymann diagrams.
    We expect this cancellation of UV
    divergences in {\em frequency} is a general feature of
    {\em non-relativistic} quantum field theories describing zero-temperature
    quantum phase transitions.

  From Eq.\ref{add1}, we can identify the two constants
  $ Z_{\alpha}, Z_{2} $ in Eq.\ref{counter} in the Minimal Subtraction
  scheme
\begin{equation}
  Z_{\alpha}=1+ \frac{e^{2}}{ 4 \pi^{2} \epsilon},~~~~
  Z_{2}=1- \frac{e^{2}}{ 12 \pi^{2} \epsilon}
\label{z2}
\end{equation}
   Note $ Z_{\alpha} \neq Z_{2} $. This is due to the {\em lack} of
   Lorentz invariance of our model.

\subsection{ The type 1 vertex}

     We turn to the evaluation of the renormalization of type 1 vertex
     in Fig.3

\begin{figure}
\epsfxsize=7 cm
\centerline{\epsffile{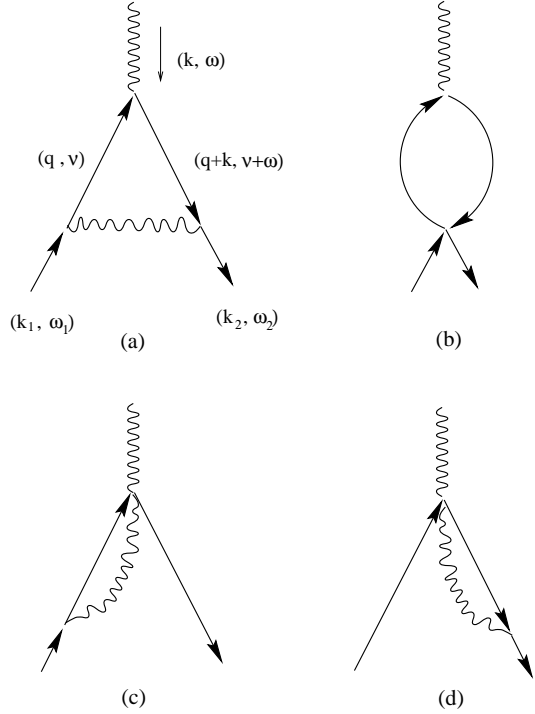}}
\caption{The one-loop diagrams of type 1 vertex  }
\label{fig: The renormalization of the type 1 vertex}
\end{figure}

     Applying the Feymann rules to Fig.3a, we have

\begin{eqnarray}
  (3a) & = & (e \mu^{\epsilon/2} )^{3}
  \int \frac{ d^{d} \vec{q}}{ (2 \pi)^{d} } \int \frac{ d \nu}{ 2 \pi}
  \frac{ (\nu+ \omega_{1})(2 \nu+\omega)(\nu+\omega+\omega_{2})  }
  { (\vec{q}- \vec{k}_{1})^{2}(q^{2}+ \nu^{2} )  
  ( (\vec{q}+ \vec{k})^{2} +( \nu+ \omega )^{2} ) }   \nonumber \\
   & = & \frac{e^{2}}{ 4 \pi^{2} \epsilon} (\omega_{1}+ \omega_{2} ) + \cdots
\end{eqnarray}

   It is easy to see Fig.3b vanishes identically and Fig.3c and Fig.3d are
\begin{eqnarray}
  (3c) &= & -\frac{e^{2}}{ 2 \pi^{2} \epsilon} \omega_{1} + \cdots
				     \nonumber  \\
  (3d) &= & -\frac{e^{2}}{ 2 \pi^{2} \epsilon} \omega_{2}  + \cdots
				     \nonumber  \\
  (3c)+(3d) &=& -\frac{e^{2}}{ 2 \pi^{2} \epsilon} (\omega_{1}+\omega_{2} ) 
  + \cdots
\end{eqnarray} 
  
    Overall, we find for the type 1 vertex
\begin{equation}
  (3a)+(3b)+(3c)+(3d)= -\frac{e^{2}}{ 4 \pi^{2} \epsilon} 
  (\omega_{1}+ \omega_{2} ) + \cdots
\end{equation}

     The constant $ Z_{1} $ in Eq.\ref{begin} can be identified
\begin{equation}
  Z_{1}= 1+\frac{e^{2}}{ 4 \pi^{2} \epsilon} 
\label{z1}
\end{equation}

       From Eqs.\ref{z2},\ref{z1}, it is evident that the Ward identity
  Eq.\ref{ward} indeed holds. The Ward identity can also be shown
  by evaluating the renormalization of the seagull vertex.
     
\subsection{ The polarization of gauge field}

       Next, we calculate the vacuum polarization graph of the $ U(1) $
  gauge field $ A_{\mu} $ in Fig.4

\begin{figure}
\epsfxsize=12 cm
\centerline{\epsffile{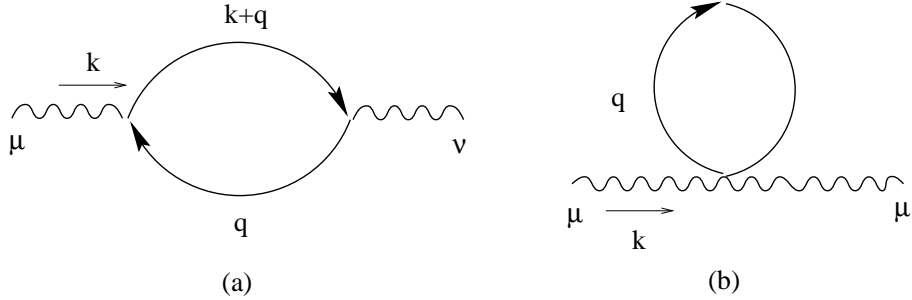}}
\caption{The vacuum polarization of gauge fields }
\label{gauge}
\end{figure}

     The fermion bubble has Lorentz invariance, therefore,
     the momenta in Fig.4 are $ D= d+1 =4- \epsilon $ dimensional
     momenta
\begin{eqnarray}
  \Pi_{\mu,\nu} ( k ) &= & e^{2} \mu^{\epsilon} N
  \int \frac{ d^{D} q}{ (2 \pi)^{D} }
  \frac{ (k+2 q)_{\mu} (k+ 2q)_{\nu} }{ q^{2} (k+q)^{2} }
  -2 e^{2} \mu^{\epsilon} N \int \frac{ d^{D} q}{ (2 \pi)^{D} }
  \frac{1}{ q^{2} }    \nonumber  \\
     & = &  -\frac{ N e^{2}}{ 24 \pi^{2} \epsilon}
     (k^{2} \delta_{\mu \nu} -k_{\mu} k_{\nu}) + \cdots
\end{eqnarray}
    where $ N $ comes from the summation over the boson suffix $ m $.

     In Eq.\ref{begin}, only the time component of the gauge field $ A_{\mu} $
     is involved. Putting $ \mu=\nu=0 $ in the above Eq. we find
\begin{equation}
  \Pi_{0,0} ( \vec{k}, \omega ) = 
  -\frac{ N e^{2}}{ 24 \pi^{2} \epsilon} k^{2} + \cdots 
\end{equation}
    where $ k $ is $ d $ dimensional space momentum.
       
    The constant $ Z_{3} $ in Eq.\ref{begin} can be identified as
\begin{equation}
  Z_{3}= 1- \frac{ N e^{2}}{  24 \pi^{2} \epsilon} 
\label{z3}
\end{equation}

\subsection{ The renormalization of the quartic term } 

    In this subsection, we compute the renormalization of the quartic term.
   
   First, let us look at the contributions from the quartic coupling 
   in Fig.5.

\begin{figure}
\epsfxsize=12 cm
\centerline{\epsffile{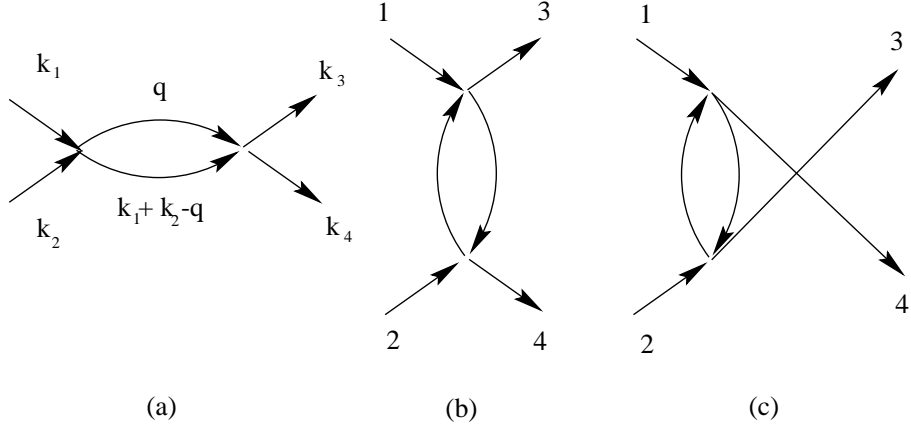}}
\caption{The renormalization from the quartic coupling }
\label{figphi4}
\end{figure}

   As in Fig.\ref{gauge}, the momenta in the above figure
   are also $ D $ dimensional momenta. 

\begin{eqnarray}
  (5a) & = & \frac{ u^{2} \mu^{ 2 \epsilon} }{2}
  \int \frac{ d^{D} q}{ (2 \pi)^{D} }
  \frac{ 1 }{ q^{2} (k_{1}+ k_{2}-q)^{2} }    \nonumber  \\
     & = & \frac{ u^{2} }{2} \frac{1}{ 8 \pi^{2} \epsilon} + \cdots
\end{eqnarray}

     Similarly, we can calculate the contributions from Fig.5b and 5c
     \cite{finite}
\begin{equation}
  (5b)= (5c) = \frac{ (N+3) u^{2} }{4} \frac{1}{ 8 \pi^{2} \epsilon} + \cdots
\end{equation}
     
     Adding all the contributions from Fig.\ref{figphi4}, we get
\begin{equation}
  (5a)+ (5b)+ (5c) = \frac{ (N+4) u^{2} } { 16 \pi^{2} \epsilon} + \cdots
\label{trick1}
\end{equation}

   Second, let us look at the contributions from the seagull term in Fig.6

\begin{figure}
\epsfxsize=10 cm
\centerline{\epsffile{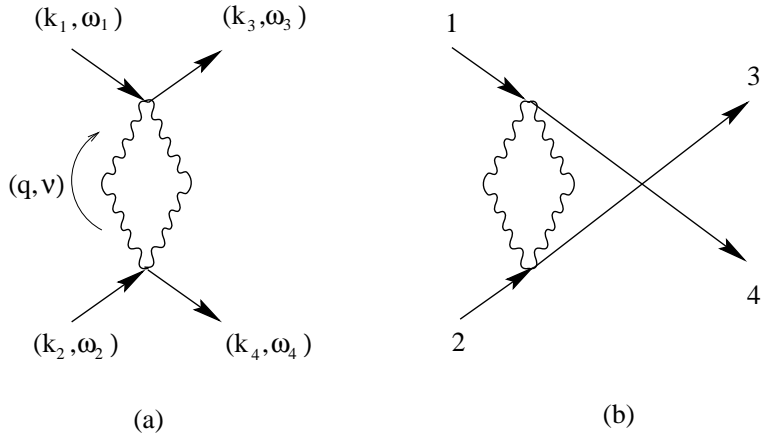}}
\caption{The renormalization from the seagull term }
\label{fig:seagull}
\end{figure}

\begin{eqnarray}
  (6a) &= & 2 (e^{2} \mu^{\epsilon})^{2}
  \int \frac{ d^{d} \vec{q}}{ (2 \pi)^{d} }
  \frac{ 1 }{ q^{2} ( q+ k_{1}-k_{3} )^{2} } \int \frac{ d \nu}{ 2 \pi}
			    \nonumber   \\
  (6b) &= & 2 (e^{2} \mu^{\epsilon})^{2}
  \int \frac{ d^{d} \vec{q}}{ (2 \pi)^{d} }
  \frac{ 1 }{ q^{2} ( q+ k_{1}-k_{4} )^{2} } \int \frac{ d \nu}{ 2 \pi}
			      \nonumber  \\
  (6a) + (6b) & = &  2 (e^{2} \mu^{\epsilon} )^{2}
  \int \frac{ d^{d} \vec{q}}{ (2 \pi)^{d} }
  \frac{ 1 }{ (\vec{q}- \vec{k}_{1})^{2}}
  ( \frac{1}{ (\vec{q}- \vec{k}_{3})^{2}}+
  \frac{1}{ (\vec{q}- \vec{k}_{4})^{2}}) \int \frac{ d \nu}{ 2 \pi}   
\label{trick2}
\end{eqnarray}

   As in Eq.\ref{self1} and \ref{self2}, we encounter the UV
   divergences in the  frequency integrals.
    We leave them alone at this moment and go ahead to calculate
    the diagrams in Fig.7

\begin{figure}
\epsfxsize=8 cm
\centerline{\epsffile{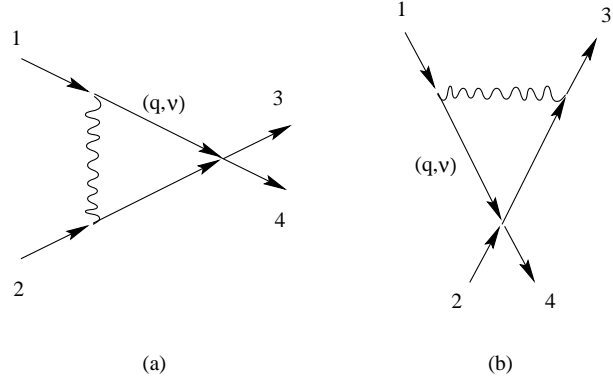}}
\caption{The renormalization from the combination of the type 1 term
+ the quartic term }
\label{fig:onephi4}
\end{figure}

   The corresponding expressions are

\begin{eqnarray}
  (7a) & = & -u (e \mu^{\epsilon/2} )^{2}
  \int \frac{ d^{d} \vec{q}}{ (2 \pi)^{d} } \int \frac{ d \nu}{ 2 \pi}
  \frac{ (\nu+ \omega_{1})(\omega_{1}+ 2\omega_{2} -\nu )  }
  { (\vec{q}- \vec{k}_{1})^{2}(q^{2}+ \nu^{2} )  
   ( ( \vec{k}_{1} + \vec{k}_{2} -\vec{q} )^{2} +
   (  \omega_{1} +\omega_{2} - \nu )^{2} ) }   \nonumber \\
     & = & \frac{ u e^{2}}{ 8 \pi^{2} \epsilon} + \cdots
                                       \nonumber   \\
  (7b) & = & -u (e \mu^{\epsilon/2} )^{2}
  \int \frac{ d^{d} \vec{q}}{ (2 \pi)^{d} } \int \frac{ d \nu}{ 2 \pi}
  \frac{ (\nu+ \omega_{1})(2 \omega_{3}- \omega_{1} +\nu )  }
  { (\vec{q}- \vec{k}_{1})^{2}(q^{2}+ \nu^{2} )  
   ( ( \vec{k}_{3} - \vec{k}_{1} +\vec{q} )^{2} +
   (  \omega_{3} -\omega_{1} + \nu )^{2} ) }   \nonumber \\
     & = & - \frac{ u e^{2}}{ 8 \pi^{2} \epsilon} + \cdots
\end{eqnarray}

    It is important to note that Fig.7a and Fig.7b have opposite signs.
   Actually, there are {\em two} diagrams in class (7a) and {\em four} diagrams
   in class (7b) corresponding to different ways to put the photon line,
   so the overall contributions are
\begin{equation}
  2(7a)+ 4 (7b)= - \frac{ u e^{2}}{ 4 \pi^{2} \epsilon} + \cdots
\label{trick3}
\end{equation}

     Let us look at the contributions from the type 1 vertex in Fig.8

\begin{figure}
\epsfxsize=12 cm
\centerline{\epsffile{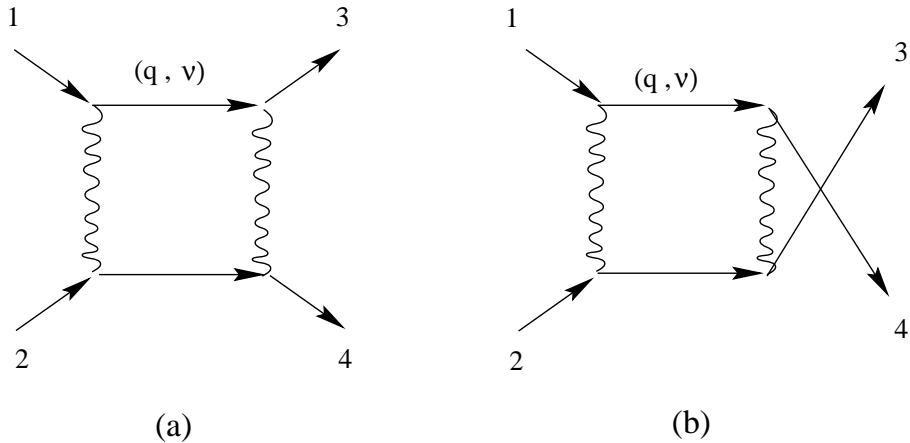}}
\caption{The renormalization from the type 1 term }
\label{one}
\end{figure}
    
     The corresponding expressions are

\begin{eqnarray}
  (8a) & = &  (e \mu^{\epsilon/2} )^{4}
  \int \frac{ d^{d} \vec{q}}{ (2 \pi)^{d} } \int \frac{ d \nu}{ 2 \pi}
  \frac{ (\nu+ \omega_{1})( \nu+ \omega_{3} )
  (\omega_{1}+ 2\omega_{2} -\nu )  (\omega_{1}+ \omega_{2} + \omega_{4} -\nu ) }
  { (\vec{q}- \vec{k}_{1})^{2} (\vec{q}- \vec{k}_{3})^{2}
  (q^{2}+ \nu^{2} ) ( ( \vec{k}_{1} + \vec{k}_{2} -\vec{q} )^{2} +
   (  \omega_{1} +\omega_{2} - \nu )^{2} ) }   \nonumber \\
     & = &  (e \mu^{\epsilon/2} )^{4}
  \int \frac{ d^{d} \vec{q}}{ (2 \pi)^{d} }
  \frac{ 1 }{ (\vec{q}- \vec{k}_{1})^{2} (\vec{q}- \vec{k}_{3})^{2}}
   \int \frac{ d \nu}{ 2 \pi}
   - \frac{ 3 e^{4}}{ 8 \pi^{2} \epsilon} + \cdots  \nonumber  \\
  (8b) & = &  (e \mu^{\epsilon/2} )^{4}
  \int \frac{ d^{d} \vec{q}}{ (2 \pi)^{d} } \int \frac{ d \nu}{ 2 \pi}
  \frac{ (\nu+ \omega_{1})( \nu+ \omega_{4} )
  (\omega_{1}+ 2\omega_{2} -\nu )  (\omega_{1}+ \omega_{2} + \omega_{3} -\nu ) }
  { (\vec{q}- \vec{k}_{1})^{2} (\vec{q}- \vec{k}_{4})^{2}
  (q^{2}+ \nu^{2} ) ( ( \vec{k}_{1} + \vec{k}_{2} -\vec{q} )^{2} +
   (  \omega_{1} +\omega_{2} - \nu )^{2} ) }   \nonumber \\
     & = &  (e \mu^{\epsilon/2} )^{4}
  \int \frac{ d^{d} \vec{q}}{ (2 \pi)^{d} }
  \frac{ 1 }{ (\vec{q}- \vec{k}_{1})^{2} (\vec{q}- \vec{k}_{4})^{2}}
   \int \frac{ d \nu}{ 2 \pi}
   - \frac{ 3 e^{4}}{ 8 \pi^{2} \epsilon} + \cdots  \nonumber  \\
\end{eqnarray}

     Actually, there are two diagrams in class (8a) and two diagrams
     in class (8b) corresponding to different ways to put the photon line,
   so the overall contributions from Fig.8 are
\begin{equation}
  2(8a)+ 2 (8b)  =  2 (e \mu^{\epsilon/2} )^{4}
  \int \frac{ d^{d} \vec{q}}{ (2 \pi)^{d} }
  \frac{ 1 }{ (\vec{q}- \vec{k}_{1})^{2}}
  ( \frac{1}{ (\vec{q}- \vec{k}_{3})^{2}}+
  \frac{1}{ (\vec{q}- \vec{k}_{4})^{2}}) \int \frac{ d \nu}{ 2 \pi}   
   -  \frac{ 3 e^{4}}{ 2 \pi^{2} \epsilon} + \cdots
\label{trick4}
\end{equation}
 
      Finally, we need to compute the diagrams in Fig.9

\begin{figure}
\epsfxsize=12 cm
\centerline{\epsffile{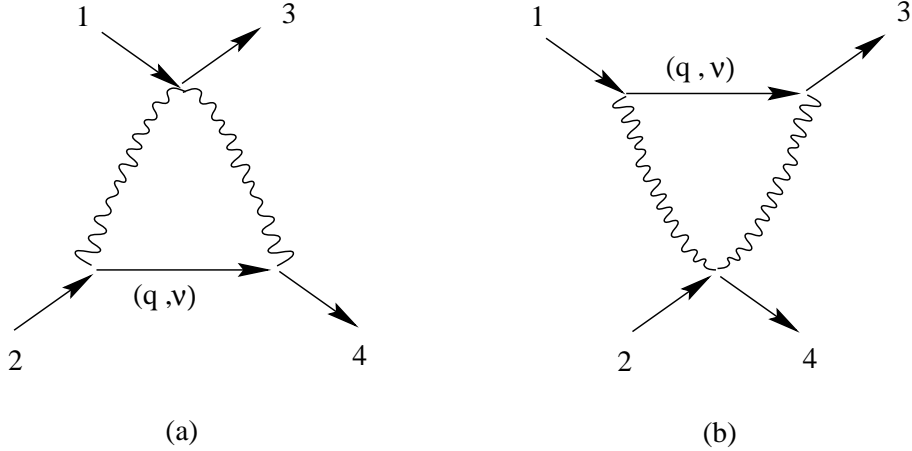}}
\caption{The renormalization from the type 1 term + the seagull term }
\label{onetwo}
\end{figure}

      The corresponding expressions are
\begin{eqnarray}
  (9a) & = & -2 (e \mu^{\epsilon/2} )^{2} e^{2} \mu^{\epsilon}
  \int \frac{ d^{d} \vec{q}}{ (2 \pi)^{d} } \int \frac{ d \nu}{ 2 \pi}
  \frac{ (\nu+ \omega_{2})( \nu+ \omega_{4} ) }
  { (\vec{q}- \vec{k}_{2})^{2} (\vec{q}- \vec{k}_{4})^{2}
  (q^{2}+ \nu^{2} ) }   \nonumber \\
     & = & -2 (e \mu^{\epsilon/2} )^{4}
  \int \frac{ d^{d} \vec{q}}{ (2 \pi)^{d} }
  \frac{ 1 }{ (\vec{q}- \vec{k}_{2})^{2} (\vec{q}- \vec{k}_{4})^{2}}
   \int \frac{ d \nu}{ 2 \pi}
   +  \frac{ e^{4}}{ 2 \pi^{2} \epsilon} + \cdots  \nonumber  \\
     (9b) & = & -2 (e \mu^{\epsilon/2} )^{4}
  \int \frac{ d^{d} \vec{q}}{ (2 \pi)^{d} }
  \frac{ 1 }{ (\vec{q}- \vec{k}_{1})^{2} (\vec{q}- \vec{k}_{3})^{2}}
   \int \frac{ d \nu}{ 2 \pi}
   +  \frac{  e^{4}}{ 2 \pi^{2} \epsilon} + \cdots 
\end{eqnarray}

    Exchanging the leg 3 with leg 4 of Fig.9a and 9b, we get another two
    diagrams (9c) and (9d) which were not shown explicitly

\begin{eqnarray}
    (9c)  & = & -2 (e \mu^{\epsilon/2} )^{4}
  \int \frac{ d^{d} \vec{q}}{ (2 \pi)^{d} }
  \frac{ 1 }{ (\vec{q}- \vec{k}_{2})^{2} (\vec{q}- \vec{k}_{3})^{2}}
   \int \frac{ d \nu}{ 2 \pi}
   +  \frac{  e^{4}}{ 2 \pi^{2} \epsilon} + \cdots  \nonumber  \\
     (9d) & = & -2 (e \mu^{\epsilon/2} )^{4}
  \int \frac{ d^{d} \vec{q}}{ (2 \pi)^{d} }
  \frac{ 1 }{ (\vec{q}- \vec{k}_{1})^{2} (\vec{q}- \vec{k}_{4})^{2}}
   \int \frac{ d \nu}{ 2 \pi}
   +  \frac{  e^{4}}{ 2 \pi^{2} \epsilon} + \cdots
\end{eqnarray}

   The over all contributions are
\begin{equation}
   (9a)+(9b)+(9c)+(9d)  =  -4 (e \mu^{\epsilon/2} )^{4}
  \int \frac{ d^{d} \vec{q}}{ (2 \pi)^{d} }
  \frac{ 1 }{ (\vec{q}- \vec{k}_{1})^{2}}
  ( \frac{1}{ (\vec{q}- \vec{k}_{3})^{2}}+
  \frac{1}{ (\vec{q}- \vec{k}_{4})^{2}}) \int \frac{ d \nu}{ 2 \pi}   
   +  \frac{ 2 e^{4}}{  \pi^{2} \epsilon} + \cdots
\label{trick5}
\end{equation}

   Adding all the contributions from Eqs. \ref{trick1},\ref{trick2}
   \ref{trick3},\ref{trick4},\ref{trick5}, we find the UV divergences
   in the frequency integrals indeed cancel and lead to the constant
   $ Z_{4} $ in Eq.\ref{begin}
\begin{equation}
Z_{4}  =    1+ \frac{ ( N+4) u }{ 16 \pi^{2} \epsilon }
  -\frac{e^{2}}{ 4 \pi^{2} \epsilon} + \frac{ e^{4} }{ 2 \pi^{2} u \epsilon}
\end{equation}
  
\subsection{ The calculation of $\beta$ function}

   In the previous subsections, we did the explicit calculations of
   the renormalization constants by considering
a direct perturbative expansion in the Coulomb fine structure constant 
$ w=e^{2} $ and the quartic coupling $ u $.
At one loop order, the values of the renormalization constants are 
summarized as
\begin{eqnarray}
Z_{\alpha} & = & 1 + \frac{w}{ 4 \pi^{2} \epsilon}   \nonumber   \\
Z_{2} & = & 1 - \frac{w}{ 12 \pi^{2} \epsilon}   \nonumber  \\
Z_{3} & =  & 1- \frac{ N w}{ 24 \pi^{2} \epsilon}    \nonumber  \\
Z_{4} & =  &  1+ \frac{ ( N+4) u }{ 16 \pi^{2} \epsilon }
  -\frac{w}{ 4 \pi^{2} \epsilon} + \frac{ w^{2} }{ 2 \pi^{2} u \epsilon}
\label{cons}
\end{eqnarray}

   Substituting the above equation into Eq.\ref{conn}, we find
   the $ \beta $ functions for $ w $ and $ u $
\begin{eqnarray}
  \beta (w) & = & -\epsilon w + \frac{ N+4}{ 24 \pi^{2} } w^{2}  
	  \nonumber  \\
  \beta (u) & = & -\epsilon u + \frac{ (N+4) u^{2} }{ 16 \pi^{2} }
      - \frac{ w u}{ 4 \pi^{2} }+ \frac{w^{2}}{ 2 \pi^{2}}
\label{beta}
\end{eqnarray}

   The fixed points are
\begin{eqnarray}
   w^{\ast} & = & \frac{ 24 \pi^{2}}{ N+4 } \epsilon    \nonumber  \\
   u^{\ast}_{\pm} & = & \frac{ 8 \pi^{2} \epsilon }{ N+4} [ 1+ \frac{6}{N+4}
   \pm \frac{\Delta}{ N+4}]
\label{fix}
\end{eqnarray}
    Where $ \Delta= \sqrt{ N^{2}-52 N-188} $.

    It is easy to see that $ \Delta $ is imaginary for $ N < N_{c}= 55.39 $.
 This $ N_{c} $ is much smaller than the corresponding critical number
 $ n_{c}=365 $ in the Scalar Electrodynamics studied in Ref. \cite{bert}.
 For $ N< N_{c} $, $ u $ is complex for both fixed points and physically
 inaccessible. The only accessible fixed points are the Gaussian
 and Heisenberg fixed points, both of which are unstable to turning
 on Coulomb interaction.  There is a runaway flow to {\em negative} value
 of $ u $ which was interpreted as a fluctuation-driven first order
 transition in Ref.\cite{bert}. For $ N > N_{c} $, these fixed points
 have real values for $ u^{\ast} $ and are physically accessible.

   In order to compare our $ \beta $ functions with those of FG,
   we put $ N=1 $ in Eq.\ref{beta}
\begin{eqnarray}
  \beta (w) & = & -\epsilon w + \frac{ 5}{ 24 \pi^{2} } w^{2}  
	  \nonumber  \\
  \beta (u) & = & -\epsilon u + \frac{ 5 u^{2} }{ 16 \pi^{2} }
      - \frac{ w u}{ 4 \pi^{2} }+ \frac{w^{2}}{ 2 \pi^{2}}
\label{N=1}
\end{eqnarray}
   
    Setting $ w= \frac{ 8 \pi^{2} }{5} w_{fg}, u=32 \pi^{2} u_{fg} $, the
    above equations become
\begin{eqnarray}
  \beta (w_{fg}) & = & -\epsilon w_{fg} + \frac{1}{3} w_{fg}^{2}  
	  \nonumber  \\
  \beta (u_{fg}) & = & -\epsilon u_{fg} + 10 u_{fg}^{2}
      - \frac{2}{5} w_{fg} u_{fg}+ \frac{1}{25} w_{fg}^{2}
\label{betafg}
\end{eqnarray}

   Putting $ \epsilon_{\sigma}=0 $ in Eq.(7) of Ref.\cite{coul}, we find
    the R. G. flow equations got by FG are different from ours.
    The $ \beta $ function should be gauge-independent at least at
    one-loop order ( see also the discussions in the appendix on this
    point).  No details being given by FG, we can only guess the possible
    mistakes made by FG: (1) the anisotropy between space and time
      (namely the dynamic exponent z ) was not treated correctly by
      FG (2) The Ward identity in the time component 
      Eq.\ref{ward} is violated in the momentum shell
      method employed by FG.
    
   The disagreement at $ \epsilon_{\sigma}=0 $  puts doubts
   of the conclusions reached by FG at Coulomb interaction case
   $ \epsilon=\epsilon_{\sigma} $. 

  Let us look at the solution right at the critical dimension $ d=3 $,
  Putting $ \epsilon=0 $ in Eq.\ref{beta} and taking the ratio of
  the two equations, we find the trajectories of the RG flow 
  satisfy the homogeneous differential equation:

\begin{equation}
  \frac{d u}{d w}= \frac{3}{2} (\frac{u}{w} )^{2}-\frac{6}{ N+4}
    \frac{u}{w} + \frac{12}{ N+4}
\label{diff}
\end{equation}

  If $ N>N_{c} $, the solution of this equation is
\begin{equation}
  C+ \ln w= \frac{N+4}{\Delta} \ln \mid \frac{ \frac{u}{w}-
  (\frac{2}{N+4} +\frac{1}{3} )-\frac{\Delta}{3 (N+4) }}
  { \frac{u}{w}-
  (\frac{2}{N+4} +\frac{1}{3} )+\frac{\Delta}{3 (N+4) }} \mid
\end{equation}
   Where $ C $ is an arbitrary constant. 
      All the trajectories flow to the origin with two marginally
      irrelevant couplings $ w $ and $ u $. At finite temperature,
      they will lead to logarithmic corrections to naive scaling functions 

  If $ N<N_{c} $, the solution of this equation is
\begin{equation}
  C+ \ln w= \frac{ 2(N+4)}{\Delta} \arctan [ \frac{3(N+4)}{\Delta}
  ( \frac{u}{w}- (\frac{2}{N+4} +\frac{1}{3} ))]
\end{equation}
      All the trajectories flow to the negative values of $ u $, indicating
   a fluctuation driven first order transition.

\subsection{ The calculation of critical exponents}

    In this subsection, we calculate the critical exponents of
    the second order transition when $ N > N_{c} $.

   From Eqs.\ref{dynamic},\ref{cons},\ref{fix},
   we find the dynamic exponent $ z $
\begin{equation}
    z= 1-\frac{4 \epsilon}{ N+4}
\end{equation}

   From Eqs.\ref{eta},\ref{cons},\ref{fix}, we find the exponent $ \eta $
\begin{equation}
   \eta= \frac{2 \epsilon}{ N+4}
\label{etano}
\end{equation}
 
   From Eqs.\ref{etaA},\ref{cons},\ref{fix}, we find the exponent $ \eta_{A} $
\begin{equation}
   \eta_{A} = \frac{N \epsilon}{ N+4}
\end{equation}
  
    The long-range interaction is modified from the form
    $\sim r^{-1+\epsilon} $ to the form $ \sim r^{-1+4 \epsilon/(N+4)} $. 

    $ Z_{\phi^{\dagger} \phi} $ can be calculated from Fig.10
\begin{figure}
\epsfxsize=12 cm
\centerline{\epsffile{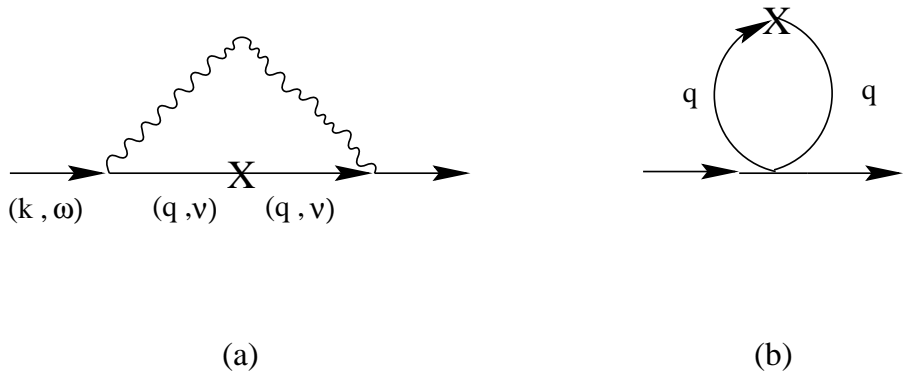}}
\caption{The cross stands for the insertion of the composite operator at
    zero momentum}
\label{fig:comp}
\end{figure}

     The corresponding expressions are

\begin{eqnarray}
  (10a) & = & e^{2} \mu^{\epsilon}
  \int \frac{ d^{d} \vec{q}}{ (2 \pi)^{d} } \int \frac{ d \nu}{ 2 \pi}
  \frac{ (\nu + \omega )^{2} }{ (\vec{q}- \vec{k})^{2}
  (q^{2}+ \nu^{2}+ m^{2} )^{2} }     \nonumber  \\
      & = & \frac{ e^{2} }{ 8 \pi^{2} \epsilon} + \cdots  
				      \nonumber  \\
   (10b) & = & -\frac{ u }{ 2} (N+1) \int \frac{ d^{D} q}{ (2 \pi)^{D}}
    \frac{ 1} { ( q^{2} + m^{2} )^{2} }    \nonumber   \\
       & = & -\frac{ u ( N+1) }{ 16 \pi^{2} \epsilon} + \cdots
\end{eqnarray}
    where we put the mass of the boson $ m $ into the boson propagators
    in order to avoid the {\em infrared} divergence.
   
      In contrast to the self energy diagrams Fig.2, there is {\em no}
   UV frequency divergence in the above equations.

   From the above equation, we can identify
\begin{equation}
    Z_{2} Z_{\phi^{\dagger} \phi}=1 +\frac{ u ( N+1) }{ 16 \pi^{2} \epsilon} 
      -\frac{ e^{2} }{ 8 \pi^{2} \epsilon}   
\end{equation}

    Substituting the value of $ Z_{2} $ in Eq.\ref{z2}, we get
\begin{equation}
   Z_{\phi^{\dagger} \phi}= 1+ \frac{u ( N+1) }{ 16 \pi^{2} \epsilon}
	 -\frac{w}{ 24 \pi^{2} \epsilon }
\end{equation}

   The correlation length exponent $ \nu $ is given by
\begin{equation}
  \nu^{-1}= 2-\frac{N+1}{ 2(N+4) } \epsilon [ \frac{N-1}{N+1}
      +\frac{6}{N+4} \pm \frac{\Delta}{N+4} ]
\end{equation}

\section{Discussions and Conclusions}

     In this paper, we provide a rather detailed derivation of RG
     analysis of Eq.\ref{fg}. We find the delicate cancellation
     of the UV {\em frequency integral} divergences, therefore, we only
     need to regulize the UV divergences in momenta integral.
     We believe that this is due to the space-independent gauge invariance
     Eq.\ref{tran} ( or the Ward identity in the time component Eq.\ref{ward} ).
     Similar cancellations also appear in the
    Dirac fermion model studied in Ref.\cite{fermion,long}.
     We expect this is a very general feature of zero temperature
     quantum critical phenomena \cite{hertz}. This feature in
     {\em non-relativistic} quantum field theory reminds us of the theorem
     in {\em relativistic} quantum field theory that
     infra-red divergences must be cancelled out in any physical
     processes.
 
     Here, we would like to make a brief comparison between the boson model
     studied in this paper and the Dirac fermion model studied in
     Ref.\cite{fermion,long}.
     In Dirac fermion model, both Coulomb and Chern-Simon
     couplings become marginal at $ d=2 $, the competition between this two
     couplings produce a line of fixed points with $ z=1 $ and
     {\em non-vanishing} renormalized Coulomb coupling. The fermion
     quartic term is irrelevant at the non-interacting fixed point and
     {\em remains } irrelevant on this $ z=1 $ line of fixed points.
     In the boson model
     studied in this paper, both the Coulomb and the boson quartic couplings
     become marginal at $ d=3 $.
     Fixing $ \sigma  $ to be 2 ( namely $ \epsilon_{\sigma} 
     \equiv 0 $ ), we performed a $ \epsilon= 3-d $
     expansion. $ \sigma \equiv 2 $ corresponds to $ 1/r $ Coulomb interaction
     at $ d=3 $, $1/r^{d-2} $ interaction at space dimension $ d $
     and {\em logarithmic } instead of $ 1/r $ interaction at $ d=2 $.
     The Coulomb coupling was found to drive the quartic coupling 
     to negative value, indicating a {\em possible} first order transition.

     Halperin, Lubensky and Ma (HLM) \cite{bert} investigated the model 
     describing the Superconducting to Normal (SN) and
     Nematic to Smectic-A transition in $ d=3 $ bulk system.
     By performing $ \epsilon=4-d $ expansion, they found
     a new stable fixed point with non-vanishing charge if the number of the
     order parameter components $ N> M_{c}=365.9/2 $, but runaway RG if
     $ N< M_{c} $ (see also the appendix). HLM interpreted this runaway RG
     flow as indicating fluctuation-driven first-order transition. Later,
     Dasgupta and Halperin(DH) stuided the model directly on a
     $ 3d $ lattice \cite{invert}. By using Monte carlo and duality arguments,
     they found the transition should be a second order transition in the
     universality class of the inverted XY-model instead of a first-order
     transition. DH's results indicate that the $ \epsilon $ expansion
     may break down at $ d=3 $.

     In this paper, we find the critical value $ N_{c} $ in the pure Coulomb
     case is much smaller than the corresponding value $ M_{c} $ in the $ 3d $
     SN transition ( more precisely $ 4D $ Scalar Electrodynamics ). This shows
     that the fluctuation is considerably reduced in the pure Coulomb case.
     We believe that it is more likely the SI transition at $ d=2 $ should
     be governed by a non-trivial fixed point with non-vanishing Coulomb
     coupling and dynamic exponent $ z=1 $

  In order to understand the effects of $ 1/r $ interaction at $ d=2 $
  near the 3D XY fixed point, two possible expansions can be used.
  One route is to fix at $ d= 2 $ and perform large $ N $ expansion.
  Unfortunately, as shown in this paper, the direct large $ N $
  expansion can only probe the physics at $ N > N_{c} $, therefore, 
  is not very useful at physical
  case $ N=1 $ in {\em sharp contrast} to the Dirac fermion case as
  discussed in Ref.\cite{fermion,long}. 
  Another route is the double expansion in $ \epsilon=3-d $ and 
  $ \epsilon_{\sigma}=2-\sigma $ performed by FG. However,
  we  believe that when {\em gauge field} fluctuations are involved,
   extrapolating to the physical case $ \epsilon=\epsilon_{\sigma}=1 $
   can only lead to misleading results.
   A method directly works at $ D=2+1 $ is needed.

   In order to make serious attempts to compare with experiments
   \cite{film,mag,junction,junction1,3d},
   disorder has to be incorporated into Eq.\ref{fg}. As shown
   in Ref.\cite{glass}, due to Griffith effects,
   even weak disorder presumably produces a gapless 
   Bose glass phase between the Mott insulator and superconducting phase.
   It is certainly  very important to study the effects of Coulomb interaction
   on the transition from superconductor to this boson glass.  It was suggested
   that with Coulomb interaction, the dynamic exponent $ z $ should be equal
   to 1 at all $ d $ and the renormalized Coulomb coupling
   is {\em finite} at the transition \cite{z=1}. This scenario was confirmed
   by extensive Monte-Carlo simulations of bosons in a 2d disordered medium 
   \cite{wal}.
   It will be welcome if we can establish
   this scenario {\em analytically} on this concrete model. 
   This scenario was indeed established
   in a clean 2d lattice model displaying quantum transitions
   from Quantum Hall state to insulating state \cite{fermion}.

\centerline{\bf Acknowledgments}
I thank B.~Halperin, S. Kivelson, A.~Millis, G. Murthy, L. Radzihovsky,
N.~Read and
S. Sachdev, Z. Tesanovic  for helpful discussions. I also thank A. M. Goldman
for pointing out Ref.\cite{3d} to me.
This work is supported by NSF Grant No. DMR-97-07701.

\appendix

\section{ The calculation of scalar electrodynamics in Feymann gauge}

     The Lagrangian studied for Scalar Electrodynamics in
     Refs.\cite{bert,chen,hi} is
\begin{eqnarray}
   {\cal L} & = & | ( \partial_{\mu}-i e A_{\mu} ) \phi_{m} |^{2} +
   + r |\phi_{m} |^{2} + \frac{u}{4} \mu^{\epsilon} |\phi_{m} |^{4}
   \nonumber  \\
  & + & \frac{1}{4} ( F_{\mu, \nu} )^{2}
   + \frac{1}{ 2 \lambda} (\partial_{\mu} A_{\mu} )^{2}
\label{start}
\end{eqnarray}
    where $ F_{\mu \nu}=\partial_{\mu} A_{\nu}-\partial_{\nu} A_{\mu} $.
 
    Eq.\ref{start} describes transitions from superconductor to normal metal
    and from a smectic-A to a nematic in liquid crystals at $ d=3 $.

     In all the previous work, the calculations were done in Landau gauge
     where $ \lambda =0 $. The advantage of The Landau gauge is
     that the divergent parts of Figs.7, 8, 9 all vanish. In this appendix,
     we do all the calculations in Feymann gauge where $ \lambda=1 $.
     In this gauge, Figs.7, 8, 9
     do make contributions, but the geometrical factors in front of
     all the integrals should be the {\em same} as those in
     the Coulomb interaction case, therefore, by comparing
     our final results with those obtained in Landau gauge, we can make
     a non-trivial check on our RG results.  Our final
     results show that (1) the $ \beta $ function is indeed gauge independent
     at least to one-loop (2) the calculations done in the Coulomb interaction
     case is correct.

The loop expansion requires counterterms to account for ultraviolet divergences
in momentum integrals; we write the counter terms as
\begin{eqnarray}
   {\cal L}_{CT} & = & (Z_{2}-1)| \partial_{\mu} \phi_{m} |^{2}
   + \frac{u}{4} (Z_{4}-1) \mu^{\epsilon} \alpha
   |\phi_{m} |^{4}  \nonumber  \\
   & - & i (Z_{1}-1) e  \mu^{\epsilon/2} A_{\mu}(
   \partial_{\mu} \phi_{m}^{\dagger} \phi_{m}-
   \phi_{m}^{\dagger} \partial_{\mu} \phi_{m})
   + (Z_{1}-1) e^{2} \mu^{\epsilon}  A^{2}_{\mu}
   \phi_{m}^{\dagger} \phi_{m}    \nonumber   \\
   & + & \frac{1}{4} (Z_{3}-1) ( F_{\mu, \nu} )^{2}
\end{eqnarray}

   Lorentz invariance requires $ Z_{\alpha}=Z_{2} $, therefore, only one
   constant $ Z_{2} $ is needed.

   Eq.\ref{start} is invariant under the gauge-transformation
\begin{equation}
 \phi \rightarrow \phi e^{i \Lambda (x) },~~~~
 A_{\mu} \rightarrow A_{\mu}- \frac{1}{e} \partial_{\mu} \Lambda (x)
\label{atran}
\end{equation}

  The Ward identity following from the gauge invariance dictates
\begin{equation}
     Z_{1} = Z_{2}
\end{equation}

   Using the identity, we relate the bare fields and couplings in
   ${\cal L}$ to the renormalized quantities by 
\begin{eqnarray}
    \phi_{m B} & = & Z_2^{1/2} \phi_m     \nonumber   \\
    A_{\mu B} & = & Z_{3}^{1/2} A_{\mu}   \nonumber   \\
    e_B & = & e \mu^{\epsilon/2} Z_{3}^{-1/2}   \nonumber  \\
    u_{B} & = & u \mu^{\epsilon} Z_{4} Z_{2}^{-1/2}   \nonumber  \\
    \lambda_{B} & = & \lambda Z_{3}
\label{aconn}
\end{eqnarray}
    
At one loop order, the values of the renormalization constants are 
\begin{eqnarray}
Z_{1} & = & 1 + \frac{w}{ 4 \pi^{2} \epsilon}   \nonumber   \\
Z_{3} & =  & 1- \frac{ N w}{ 24 \pi^{2} \epsilon}    \nonumber  \\
Z_{4} & =  &  1+ \frac{ ( N+4) u }{ 16 \pi^{2} \epsilon }
  -\frac{w}{ 4 \pi^{2} \epsilon} + \frac{3 w^{2} }{ 2 \pi^{2} u \epsilon}
				 \nonumber  \\
Z_{\phi^{\dagger} \phi } & = & 1 + \frac{ (N+1) u }{ 16 \pi^{2} \epsilon}
	    -\frac{ 3 w}{ 8 \pi^{2} \epsilon} 
\label{acons}
\end{eqnarray}

   Substituting the above equation into Eq.\ref{aconn}, we find
\begin{eqnarray}
  \beta (w) & = & -\epsilon w + \frac{ N}{ 24 \pi^{2} } w^{2}  
	  \nonumber  \\
  \beta (u) & = & -\epsilon u + \frac{ (N+4) u^{2} }{ 16 \pi^{2} }
      - \frac{3 w u}{ 4 \pi^{2} }+ \frac{ 3 w^{2}}{ 2 \pi^{2}}
\label{abeta}
\end{eqnarray}
   
      Eqs.\ref{acons},\ref{abeta} should be compared with the corresponding
      Eqs.\ref{cons},\ref{beta} in Coulomb interaction case.
      The differences should be noted.

      After proper scalings of $ w $ and $ u $, we find Eq.\ref{abeta}
      is the same as that derived in Landau gauge where $\lambda=0 $.

       The critical exponents are
\begin{eqnarray}
   \eta_{F} & = & -\frac{6}{N} \epsilon    \nonumber   \\
   \eta_{A} & = &  \epsilon    \nonumber   \\
   \nu^{-1} & = & 2- \frac{\epsilon}{2(N+4)} [ (N+1)-\frac{54}{N}
    \pm \frac{N+1}{N} \Delta_{a} ]
\end{eqnarray}
   Where $ \Delta_{a}= \sqrt{ N^{2}-180N-540 } $.

   It is easy to see that $ \eta $ {\em differs} from the value computed
   in Landau gauge ( $ \eta_{L}= -\frac{9}{N} \epsilon $ ).
   It is {\em not} expected to be, because the
   correlation function of $ \phi $ is not gauge invariant anyway.
   What is the physical meaning of $ \eta $ is not clear. However,
   the $ \eta $ in Eq.\ref{etano} calculated in Coulomb interaction case is
   well-defined. The reader is encouraged to read Ref.\cite{fermion,long}
   for the similar discussions in Dirac fermion case.

   $ \eta_{A}, \nu $ are the same as those computed in Landau gauge,
   because they are gauge invariant quantities.
    $ \eta_{A}=\epsilon $ indicates that the Coulomb interaction {\em remains}
    $ 1/r $ to order $ \epsilon $ at $ d=3-\epsilon $ in contrast to the pure
    Coulomb interaction case. Actually, it was shown that the
    equation $ \eta_{A} = \epsilon $ is {\em exact} \cite{igor,tes}.
   The critical
   number which divides the second order phase transition and
   the possible fluctuation-driven first order transition is $ M_{c}=365.9/2 $
   which is of course gauge-independent.


\begin{references}
\bibitem{omit} In the following, for the simplicity of notation, we
    replace $ e^{*} $ by $ e $.
\bibitem{finite} Actually, the equality only holds up to finite terms.
   In order to get the right prefactor $ N+3 $, the faithful representation
   of the quartic term in Fig.1c should be used.
\end{references}
\end{document}